\documentclass[11pt]{article}
\usepackage{amssymb}
\usepackage[dvips]{graphicx}

\renewcommand{\@}[1]{\sqrt{#1}}

\renewcommand{\le}[1]{\label{#1}\end{eqnarray}}
\newcommand{\be}{\begin{equation}}
\newcommand{\ee}{\end{equation}}
\newcommand{\bea}{\begin{eqnarray}}
\newcommand{\eea}{\end{eqnarray}}

\def\ffract#1#2{\raise .35 em\hbox{$\scriptstyle#1$}\kern-.25em/
\kern-.2em\lower .22 em \hbox{$\scriptstyle#2$}}

\begin{document}

\begin{center}
\vskip 2.5truecm {\Large \textbf{Holographic Views of the World}}\\
\vskip.5cm
{\Large On the Occasion of Gerard 't Hooft's 60th Birthday}
\end{center}

{\it This note has been written on the occasion of Gerard 't Hooft's 60th birthday celebration. It is not a technical
paper but just a collection of discussions that the three authors had with Gerard, mostly
in relation with the idea of holography and aspects of quantum black hole physics.
Leonard Susskind (Professor of
Physics at the University of Stanford)
is a leader in various areas of Theoretical Physics and, among many fundamental results, he
 has contributed to our understanding of black hole physics. He and Gerard discovered what Lenny
coined the ``holographic principle''. Giovanni Arcioni (currently a postdoc at the Hebrew University
 of Jerusalem) did a postdoc in Utrecht and spent 
 most of his Ph.D. period there. Sebastian de Haro (currently a postdoc at the Albert Einstein Institute in Potsdam)
 was Gerard's master's and
 Ph.D.~student. }

\vskip 2truecm





\section*{The Holographic Principle}

It is a real pleasure to contribute to the ``Liber Amicorum" published on the occasion of the Gerard60 Fest.
Gerard is not only one of the greatest living physicists; his ideas continue to be a source of inspiration in many fields
of physics. We will recall here some of the discussions we had with Gerard, and in addition some episodes which
occurred to all of us during these meetings. The Holographic Principle originated from discussions
between Gerard and Lenny. Below is Lenny's account of his recollections of those years. It will be followed by the
contribution of the other two younger authors on some personal experiences with Gerard.\\

Gerard and I had been talking on and off about black holes for a number of years by 1994. He and I seemed to be the only two people who 
were completely convinced that the basic quantum laws of information and entropy must be respected by black holes. Although I am sure we agreed
about the substantive issues, we tended to think a little differently. Gerard wanted to think
 about it from an $S$-matrix point of view like in
quantum field theory. He wanted to construct a unitary $S$-matrix that would evolve an in-going state to an out-going state on the horizon of a
black hole. I certainly agreed that an $S$-matrix should exist but it seemed to me  hopeless to actually compute it.  I thought that
trying to construct an $S$-matrix would be a lot harder than discovering the underlying microstructure. I had
formulated the idea of Black Hole Complementarity which stated that from the outside perspective, the (stretched) horizon of a black hole is
composed of microscopic degrees of freedom that absorb, thermalize, and re-emit all information. But I had also argued that from the infalling
point of view, the horizon was just empty space with no special properties. Think of an observer
 in a free falling elevator: as long as the
elevator is freely falling, and up till the point when it hits the ground, she won't be able to tell the difference between the laws of physics inside the small
elevator and those inside a space-ship out in space. So will, for an observer who is freely falling into a black hole, and up till the point when
she is crushed by tidal forces or absorbed in the singularity, the physics around her be the physics of empty space. Yet we know that for an
observer who stays outside or is trying to escape from the black hole -- like in an accelerator that is going up --, the region near the horizon is strongly
gravitating and in fact it has membrane-like properties like an electric surface resistivity of 377 ohms and viscosity. I argued that the
discrepancy of the two different descriptions is only apparent -- only in the case that we think in terms of some superobserver, who somehow has
access to both the freely falling and the accelerated system near the black hole, do we get any contradictions. That such a description should
be precluded is what I called Black Hole Complementarity. Like in quantum mechanics, where we can't measure both position and momentum at the same
time without disturbing the system, we can't measure both the inside and the outside of the black hole without using signals of an energy
of the order of the Planck scale. This way Black Hole Complementarity argues that the two seemingly contradictory views can be reconciled, if we just agree on which
observable we decide to measure.\\

At Stanford my younger collaborators, Larus Thorlacius, John Uglum,  and I had repeatedly marvelled over the fact that all the information that
the black hole can hold is somehow coded on the two dimensional horizon in Planckian pixels. In the Stanford physics department at that time
there was a holographic display of a very pretty flirtatious girl that we liked to look at. The idea that the horizon was like a hologram was a
sort of half-joke that we used as an excuse to look at the display.\\

In 1994 my wife Anne and I visited Gerard in Utrecht for about 3 weeks. It was an unbelievably hot summer and it was hard to concentrate on
physics in the heat. Gerard was an excellent host, acting as a tour guide for the local Dutch countryside. Anne and I have extremely fond
memories of the visit apart from the deadly heat. But temperature not withstanding, Gerard and I (I think Erik or Herman Verlinde
 also participated) did
get some time to talk about physics. Mostly he talked about his $S$-matrix ideas while I tried to convince him that the right thing was to think
about the horizon degrees of freedom and how string theory might explain them.\\

But there was one discussion very near the end of our visit during which Gerard mumbled that if you could measure everything on the boundary of a
region with Planckian resolution, you could reconstruct everything inside. Bang! The word hologram exploded in my head but I don't think either
of us explicitly used it. Gerard mentioned something about a paper he wrote. I looked for it but I couldn't find it. What had fooled me was his
use of the term ``dimensional reduction" which is used in quantum field theory, but means something else.\\

On my way home I realized that the key issue was how to argue that the maximum entropy in a region was the area. I think Gerard must have said
that but I don't remember. I followed a similar path to what he had done in his paper \cite{dimred} (which I didn't finally see for about a year). First
consider a free gas of radiation in a box. Entropy is a measure for the number of microstates of the gas inside the box. Now basic thermodynamics tells us
that the maximum entropy of the gas is proportional to the volume of the box, in other words the number of microstates inside the box is bounded
by the volume. The entropy of a black hole, on the other hand, is known to grow like the area of the horizon surrounding it, which is a smaller
quantity. Now suppose a region of space inside the volume $V$ was found to have an entropy in excess of the entropy of a black hole just big
enough to fit in $V$ but with smaller energy. By throwing in additional matter such black hole could be formed. Since the entropy of the black
hole would be smaller than the original entropy the second law of thermodynamics
would be violated. So I began to realize that the black hole
process together with the 2nd law provided a rigorous bound on the entropy
 that any system can have--the holographic bound, the maximum entropy inside a region is bounded by
the area and not by the volume of that region. By the time we arrived in San Francisco I had hand written
the paper. The only problem was that I didn't know how to reference Gerard's paper.

When I got home I looked for something with an appropriate title but I couldn't find it. As I said, dimensional reduction meant dimensional
reduction. So I ultimately referenced the closest thing I could find which turned out to be the wrong reference.
When I finally found the reference I was glad to see that I had correctly attributed the right ideas to Gerard. Up till that point I was guessing
what he wrote.

\section*{Some personal recollections}

As explained above, the holographic principle was formulated by Gerard
as an attempt at understanding the physics of quantum black holes and at
resolving some basic but difficult paradoxes that arise there, like information loss. The
outcome of Gerard's own investigations on this goes under the name of the {\it $S$-matrix
Ansatz} for black holes. Two of us, G.A. and SdH,  worked on the $S$-matrix while they were
in Utrecht. We review here, in a very informal way, some of the things we learned
from the discussions with Gerard, emphasizing his most frequent
remarks. In addition, we would also like to mention a few fond memories from
these meetings with Gerard; they reveal not
only his way of working but also his personality. Some of them could
be classified as
curiosities and we hope that they will be enjoyable for the reader as well.\\

So let's go back to our undergraduate period (around 1997,
just before the discovery of AdS/CFT correspondence). We can quite easily
reconstruct some of Gerard's first remarks to us 
concerning quantum black holes. Here they go:

\begin{description}

    \item $\bullet$ All processes in Nature, including therefore the
 evaporation process of the quantum black hole, should be
    described by a {\it unitary} $S$-matrix with in and out
    asymptotic states.
    \item $\bullet$ Black holes of the Planck scale should be indistinguishable
    from elementary particles.
    \item $\bullet$ The four-dimensional Schwarzschild black hole is the first challenge
and at the same time an excellent arena to test all ideas and
    models for quantum black holes.
    \item $\bullet$ Gravitational and Standard Model interactions
    near the black hole horizon are not negligible; moreover, particles can get transplanckian energies and these
    modes cannot be neglected but give non-trivial
    backreaction effects to be incorporated in the $S$-matrix.

\end{description}

The first point above sounded quite natural and conservative. We had
just finished our undergraduate studies a few months before,
therefore the requirement of the existence of a unitary $S$-matrix was the natural way to reconcile
general relativity with what we had learned about quantum field theory. Things are not so simple, though, for the problems under
investigation. In particular, we later realized that one the most
delicate issues is how to split the $S$-matrix operator into an in-, out-part, and a non-trivial
horizon piece. Although one might be able to handle small fluctuations around a large 
black hole, how was one supposed to deal with small black holes, precisely of the
type that should be indistinguishable from elementary particles?
This was his second point, which did seem very innovative to us: even without
invoking renormalization group arguments for the different behavior
of Yang-Mills and general relativity theories at short distances, this point of view deviates
from what one usually learns in standard General Relativity courses.
Gerard, in particular, repeated several times that the coupling
constant increases with decreasing distance scale which means that
the smaller the distance scale, the stronger the influence of black
holes; at Planckian scale it should be impossible to disentangle
black holes from elementary particles. The third point sounded natural but we realized
that most progress in microstate counting of black holes
had been done in very specific cases like the (near)-extremal black hole,
and using much stronger assumptions like supersymmetry. The
 last point above again represents a peculiar
ingredient of Gerard's $S$-matrix. The inclusion of transplanckian
modes raises questions concerning non-locality. In
addition the horizon becomes a sort of membrane which stores and
transmits information. This is reminiscent of Lenny's stretched
horizon scenario (a sort of quantum evolutum of the classical
membrane picture of General Relativity) to implement the Holographic
principle. Both of us, by the way, were reading with great interest Lenny's paper
on the subject, starting from his seminal paper \cite{Susskind:1994vu}. However, as he
explains above, trying to construct an $S$-matrix would be harder than discovering the underlying
microstructure.\\

These are the first couple of things  Gerard told us when we started to
work on his $S$-matrix. In further detail, we would like to list 
below specific aspects of the $S$-matrix approach which we
discussed with him:

\begin{itemize}

    \item  In the $S$-matrix approach, when going beyond the eikonal regime,
which means, including the transverse components of the
    gravitational field (i.e. the transverse coordinates in a
    light-cone framework), one should see the emergence of a
    discrete structure on the horizon. The explicit inclusion of the
    transverse component looks like one of the most difficult parts
    of the whole $S$-matrix approach.
    \item The non commutative algebras of ingoing and outgoing
    operators. We discussed whether a
    covariant version of these algebras, once the transverse part of
    the gravitational field is included, could resemble a
    Virasoro algebra or something similar from which one could
    eventually infer a central charge and out of it compute the black
    hole entropy.
    \item The inclusion of fermions in the $S$-matrix approach.
    This is still an unsolved problem. In general if a bulk field has spin $s$, 
    the holographic field on the horizon should have spin $s-1$ and
    this looks problematic for fermions.

\end{itemize}

Gerard always insisted on the importance of {\it deriving} rather than
postulating things, and of clearly stating the assumptions that one is making
rather than hiding them in the formalism.
Another thing which we both remember is that Gerard pointed out
several problems to solve but he never made a hierarchy indicating
which one is more difficult. Sometimes he simply remarked ``It is
difficult to go on". This answer may sound obvious but it is not if
we once again keep in mind that all the steps have
to be derived and not postulated.\\

We also worked on the 2+1 version of the
$S$-matrix that Gerard also worked on. In this case a very intriguing
horizon algebra could be derived which was
remarkably covariant, but with a non-compact group, which is a
problem if one wants to get a finite bound for the black hole
entropy. Discussing with him we realized that Gerard was re-deriving
the representation theory of SO(2,1), without borrowing it from
standard textbooks. This is another quite amazing thing that we
noticed: he derives the mathematics
that he needs, without reading existing literature.\\

Recently he has been trying to re-derive quantum mechanics: from what
we understand, he thinks that classical mechanics in the presence of
dissipation and non linearities should be the arena from which
quantum mechanics comes out. There should be a reduction of the
initial large number of degrees of freedom which are now organized
into what he defines as ``equivalence classes" which live on a Hilbert
space. Whether this view on quantum mechanics is true or not time will tell, but 
it seems that once again Gerard is grasping some deep aspect of 
nature here. We discussed some of these things with him, even though it
sounds like a very hard conceptual problem --one that not much progress can be
made on for the moment-- and we cannot say
we completely understand the picture he has in mind.
These discussions, however, have
been very stimulating. It is interesting to see how
Gerard, when trying to solve such a difficult problem, starts from
very simple examples which encapsulate most of the physical
aspects of the problem and then starts to construct his new theory. These examples
seem innocent but are usually very deep. Also Lenny constructed several
Gedanken experiments that illustrate the holographic principle. That ``a picture is worth
a thousand words'' certainly deserves to be rephrased here as ``a picture is worth $N$ formulas''
(where $N$ is obviously large...).\\

It would take too long to describe the discussions with Gerard on
the papers on the subject that we wrote independently
\cite{Arcioni:2004wc}\cite{deHaro:1998tj}\cite{SdH1}. We mention
here the benefit from the discussions with him on our joint paper
\cite{Arcioni:2001my} (in collaboration with Martin O'Loughlin, a
postdoc at the Spinoza Institute at that time). There we showed that for
an eikonal limit of gravity in a space-time of any dimension with a
non-vanishing cosmological constant, the Einstein -- Hilbert action
reduces to a boundary action. We stressed in particular the role of
the off-diagonal Einstein action in removing the bulk part of the action.
The crucial input which came from Gerard is that he did not believe
in a gauge where one has a 2+2 splitting of the metric with
vanishing off-diagonal blocks. After long discussions, we were convinced that he was indeed right.
This completely changed the way the paper looked like in the end, since we were forced to use a completely
different method. When we had our discussions Gerard always preferred to fix an
appointment. This is a common ``procedure" let's say. For a quick
question you can jump into his office at any time, but for a real
discussion he prefers to fix an appointment in his electronic
agenda. What we noticed, in general, is that when the discussion
starts to interest him he goes to the blackboard and starts to
write. In general he does not write many formulas, but draws pictures to
illustrate the process. Perhaps one way to characterize the way he
works is that he does not start from a theory, but from a
picture or a physical situation. Then he asks the questions he is
interested in and looks for the right variables, and the right
theory (which he usually
ends up constructing from scratch...).\\


Let's describe these discussions a bit more: usually Gerard does
not like very general stories and his explanations are brief
and to-the-point. In our experience, the best way to  profit
from a conversation with him is to ask  precise and detailed
questions. He does not like arguments that are too philosophical or general claims, but goes very quickly to
the relevant points. On some occasions, however, he
starts talking really fast and tells you lots of things about a specific
problem. 
Actually, in addition to very deep intuition and
mastering of the technical details, many times we had the impression
that Gerard ``thinks differently" from the conventional way.\\


In general all these discussions are really enjoyable and compared
to what he knows and is able to do we can definitively say that
Gerard is modest as a person. You never feel a barrier or get the sensation that he is looking down on
you. Sometimes he admitted that nowadays it is
difficult for our generation to make remarkable progress. Still, Gerard 
stimulates you to tackle the important questions, whether they are difficult or not, and 
suggests different strategies. No matter what the subject is, he usually asks the 
questions that help you rethink what you are doing from the very beginning and often bring 
the problem in an entirely different light.\\

Here is a specific situation which one of us (G.A) remembers in
connection with the fact that Gerard does not like to behave like a
``star" but looks absolutely normal and...well curious of details as
well. It happened that one day he jumped into the office I was
sharing with one of his Ph.D. students, Stephan Nobbehuis (who
recently graduated). They spoke in Dutch, and I guessed that Gerard had
started by asking him about any progress on his thesis work, a
pretty  standard question that a supervisor makes to his Ph.D.
students. What Stephan showed him was not a series of computations,
but a power-point presentation where a funny animation described
the Higgs mechanism. Stephan was showing a nice animation where the
rolling down to a vev of the potential was accompanied by an
``explosive" sound (a little bit disturbing I have to say; this
animation was based on the famous story of the poor coyote Willy
who falls down trying to get the roadrunner). Gerard seemed
to enjoy himself (quite a lot), went out fast and came back in a
couple of minutes; they tried again the animation and this time the
sound was modified (it was definitively more melodical). After this
they both looked happy, they quickly exchanged a few words and
then Gerard returned to his office. Therefore, I had just seen a
Nobel Prize winner dedicating 10-15 minutes to improve the animation of the
symmetry breaking. This episode and others showed us that Gerard
is having a lot of fun doing physics; it is
not a business where you just have to publish, but there is room
also for fun. [I wanted to see the animation as well at that point
and I kept on asking Stephan to show it to me occasionally; I
apologize to him for this...].\\

SdH recalls some anecdotes from his undergraduate period with
Gerard. Many of the undergraduates had a somewhat cliche-picture of
Gerard as the absent-minded professor. Perhaps it wasn't totally
ungrounded. At the beginning of my period as a master's student
when he was my supervisor, I would walk into his
office and he would ask me who I was. I would say my same and add ``I am
doing a master's thesis on black holes" and then he would remember. Of
course this absent-mindedness was only an appearance; it may be true
that Gerard sometimes forgets things, but he really cares
about the people around him.\\

We would like to close by mentioning a more general issue: what does
Gerard think about string theory? He has
 answered that question clearly enough in his popular science book
  ``In Search of the Ultimate Building Blocks" where, among other things, he
   says that, was it not that so many people were already doing string theory,
he would probably have done it himself. In fact, his quest for black holes has
always been intimately connected to the search for a string, or membrane,
 description of the evaporation of a black hole. Already in his seminal paper
 ``Graviton Dominance in Ultra-High-Energy Scattering'', which has been the
  basis of his considerations on the relevance of the eikonal approximation
    and scattering near a black hole, he  found that the scattering
    amplitude of two particles at Planckian center of mass energy
    has a remarkable resemblance with the Veneziano amplitude, due to the fact that the dynamics basically
    takes place on a two-dimensional surface. One could say that this was also a first indication of the
     holographic properties of gravity. We do have an anecdote to tell about this. I (SdH) remember that
      the first time that Giovanni visited in Utrecht (we were both starting our PhD's), we
      went into Gerard's office to discuss about something I can't remember, but it must have
       been about black holes. Giovanni explained what he was working on, but as soon as he said the
        word ``supersymmetry", Gerard's attention seemed to drop. Once we went back to Planckian scattering, Gerard was right on again.
         But, in fact, Gerard has always followed the developments in string theory. During my PhD,
         he once asked me what I was working on. When I said I was
          working on string theory, he smiled and mumbled something like ``I can't keep my students from doing
          that".\\

\section*{...not a last word...}

Gerard's unique contribution to physics has been reviewed by many experts at this marvelous conference. Here we just wish
to thank him for the beautiful interaction and stimulus
 over the years, and wish him a very happy 60th birthday!\\

Lenny Susskind, Giovanni Arcioni, Sebastian de Haro


\begin{thebibliography}{123}




\bibitem{Susskind:1994vu}
  L.~Susskind,
  ``The World as a hologram,''
  J.\ Math.\ Phys.\  {\bf 36}, 6377 (1995)
  [arXiv:hep-th/9409089].

\bibitem{Bigatti:1999dp}
  D.~Bigatti and L.~Susskind,
  ``TASI lectures on the holographic principle,''
  arXiv:hep-th/0002044.

\bibitem{'tHooft:1996tq}
  G.~'t Hooft,
  ``The scattering matrix approach for the quantum black hole: An overview,''
  Int.\ J.\ Mod.\ Phys.\ A {\bf 11}, 4623 (1996)
  [arXiv:gr-qc/9607022].


\bibitem{Arcioni:2004wc}
  G.~Arcioni,
  ``On 't Hooft's $S$-matrix ansatz for quantum black holes,''
  JHEP {\bf 0410}, 032 (2004)
  [arXiv:hep-th/0408005].

\bibitem{Arcioni:2001my}
  G.~Arcioni, S.~de Haro and M.~O'Loughlin,
  ``Boundary description of Planckian scattering in curved spacetimes,''
  JHEP {\bf 0107}, 035 (2001)
  [arXiv:hep-th/0104039].
\bibitem{deHaro:1998tj}
  S.~de Haro,
  ``Planckian scattering and black holes,''
  JHEP {\bf 9810}, 023 (1998)
  [arXiv:gr-qc/9806028].

\bibitem{SdH1}
  S.~de Haro,
  ``Noncommutative black hole algebra and string theory from gravity,''
  Class.\ Quant.\ Grav.\  {\bf 15} (1998) 519
  [arXiv:gr-qc/9707042].

\bibitem{dimred}
  G.~'t Hooft,
  ``Dimensional reduction in quantum gravity,''
  arXiv:gr-qc/9310026.



\end{thebibliography}
\end{document}